\documentclass[review]{elsarticle}

\pdfoutput=1
\usepackage{hyperref}
\usepackage{multirow}
\usepackage{natbib}
\usepackage{mathptmx}
\usepackage{mathrsfs}
\usepackage{mathtools}
\usepackage{epsf}
\usepackage{epsfig}
\usepackage{epstopdf}
\usepackage{graphicx}% Include figure files
\usepackage{dcolumn}% Align table columns on decimal point
\usepackage{bm}
\usepackage[toc,page]{appendix}
\usepackage[latin1]{inputenc}
\begin{document}

\begin{frontmatter}

\title{Study of single crystalline SrAgSb and SrAuSb semimetals}
%\tnotetext[mytitlenote]{Fully documented templates are available in the elsarticle package on \href{http://www.ctan.org/tex-archive/macros/latex/contrib/elsarticle}{CTAN}.}:
\author{P. Devi\corref{mycorrespondingauthor}}
%\ead{phyparul@gmail.com}
\address{Ames Laboratory, U.S. Department of Energy, Iowa State University, Ames, Iowa 50011, USA}

\author{Lin-Lin Wang}
\address{Ames Laboratory, U.S. Department of Energy, Iowa State University, Ames, Iowa 50011, USA}

\author{Caiden Abel}
\address{Department of Physics and Astronomy, Iowa State University, Ames, Iowa 50011, USA}

\author{Sergey L. Bud'ko}
\address{Ames Laboratory, U.S. Department of Energy, Iowa State University, Ames, Iowa 50011, USA}
\address{Department of Physics and Astronomy, Iowa State University, Ames, Iowa 50011, USA}

\author{Paul C. Canfield}
\address{Ames Laboratory, U.S. Department of Energy, Iowa State University, Ames, Iowa 50011, USA}
\address{Department of Physics and Astronomy, Iowa State University, Ames, Iowa 50011, USA}

\begin{abstract}
Given renewed interest in the electronic properties of semimetallic compounds with varying degrees of spin orbit coupling we have grown single crystals of SrAgSb and SrAuSb, measured their temperature and field dependent electrical resistivity and magnetization and performed density functional theory (DFT) band structure calculations. Magnetization measurements are consistent with a diamagnetic host with a small amount of local
moment bearing impurities. Although the residual resistivity ratio (RRR) for all samples studied was relatively low, ranging between 2.4 and 3.4, the compounds had non-saturating magnetoresistance (MR), reaching values of $\sim$ 17\% and $\sim$ 70\% at 4 K and 9 T for SrAgSb and SrAuSb respectively. Band structure calculations, using the experimentally determined Wyckoff positions for the Sr, Ag/Au, and Sb atoms, show that whereas SrAgSb is a topologically trivial, but compensated, semimetal; SrAuSb is a topologically non-trivial, Dirac semimetal.
\end{abstract}

\end{frontmatter}

\section{INTRODUCTION}
Compounds with novel electronic band structures are of broad interest since it can lead to unique states of matter, often reflected in unusual electronic transport properties \cite{Hasan2010}. Recently, researchers proposed and experimentally investigated many new semimetallic phases, which host non trivial band topology with vanishing densities of states near the Fermi energy. In general, topological semimetals can be classified into three different types: a Dirac semimetal, a Weyl semimetal and a nodal-line semimetal \cite{Chen2009, Jo2017}. Among all the topological semimetal compounds, Cd$_{3}$As$_{2}$ and Na$_{3}$Bi were first theoretically predicted and experimentally verified to host Dirac fermions, whereas Weyl fermions were identified in the family of materials $\textit{TX}$ ($\textit{T}$ = Ta, Nb; $\textit{X}$ = As, P) \cite{Armitage2018, Schoop2018, Shekhar2015}. These materials might be considered for technological applications because of unique transport responses such as large MR, high electronic mobility, quantum Hall effect (QHE) and so on \cite{Armitage2018, Shekhar2015, Leahy2018}. \\

The MR gives information about the characteristics of Fermi surface. The large MR in topological non-trivial materials can be understood from several mechanisms and still open to debate \cite{Hasan2010, Armitage2018, Shekhar2015, Leahy2018}, making these materials of great interest in both fundamental research and device applications \cite{Salamon2001, Wang2012}. Therefore, prediction and/or experimental discovery of new materials with topologically non-trivial band structure are of growing interest. \\

We have focused our search for alkaline earth metal based Sr- compounds i.e. SrAgSb and SrAuSb. These compounds crystallize in the BeZrSi-type structure, an ordered superstructure variant of the AlB$_{2}$ structure with planar $\textit{T$_{3}$Sb$_{3}$}$ hexagons which are rotated by 60$^{\circ}$ in every other layer. Previous studies on magnetic analogues of these compounds show that they crystallize in hexagonal structure with the centrosymmetric space group \textit{P}6$_{3}$/mmc \cite{Tomuschat1981, Tomuschat1984, Mishra2011}. The band structure calculations on these compounds (using the same space group but different from experimentally observed Wyckoff sites) suggest topological non-trivial features for both compounds SrAgSb and SrAuSb, as listed in the topological material databases  \cite{Tang2019, Zhang2019}. \\ 
 
In this paper, we report the single crystal growth, structure characterization, the band structure calculations, and magnetization and magnetotransport properties of SrAgSb and SrAuSb. Our x-ray diffraction (XRD) analysis of lattice parameters and Wyckoff sites is consistent with the literature \cite{Mishra2011, Merlo1990}. The band structure calculations show that SrAgSb is topologically trivial and SrAuSb is topologically non-trivial. Temperature and field dependent magnetization measurements are consistent with a diamagnetic host, having small amount of magnetic impurity. The transport measurement shows that both compounds exhibit low RRR with similar residual resistivity values and a non saturating MR. \\ 

\section{EXPERIMENTAL AND COMPUTATIONAL METHODS}
Single crystals of SrAgSb and SrAuSb were grown out of Sr-Ag-Sb and Sr-Au-Sb ternary melts. We put elements with an initial stoichiometry according to the molar ratio of Sr:Ag:Sb = 1:18:9; Sr:Au:Sb = 1:18:9 into a fritted alumina crucible [CCS], and then sealed the crucible into an amorphous silica tube under partial Ar atmosphere \cite{Canfield2016, Canfield2020}. This initial stoichiometry was chosen on the basis of the binary phase diagrams of Ag-Sb and Au-Sb, so that the Ag-Sb and Au-Sb ratios correspond to their binary eutectics. For growing SrAgSb crystals, the ampoule was heated up to 1050$^{\circ}$C over 5 hours, held there for 5 hours and then slowly cooled down to 800$^{\circ}$C over 99 hours, and then finally decanted using a centrifuge. Single crystals of SrAuSb were grown by heating the ampoule up to 1000$^{\circ}$C over 5 hours, holding it at the temperature for 5 hours and slowly cooling to 700$^{\circ}$C over 75 hours. Slow cooling rate was kept lower for Ag based single crystals to avoid dendrite kind of morphology \cite{Canfield2020}. The fritted crucible allowed for the clean separation of crystals from the Ag-Sb and Au-Sb flux. The single crystalline samples have a plate-like shape with the dimensions up to $\sim$ 3$\times$3$\times$0.5 mm$^{3}$ and the as-grown surface is the ab-plane.\\ 

Band structure with spin-orbit coupling (SOC) in density functional theory\cite{Hohenberg1964, Kohn1965} (DFT) has been calculated with PBE \cite{Perdew1996} exchange-correlation functional, a plane-wave basis set and projected augmented wave method as implented in VASP \cite{Kresse1996, Kresse1996a}. Experimental lattice constants \textit{and} Wyckoff sites have been used with a Monkhorst-Pack \cite{Monkhorst1976} (11$\times$11$\times$6) k-point mesh including the $\Gamma$ point and a kinetic energy cutoff of 250 eV.\\

To acquire XRD data at room temperature, a Rigaku Miniflex II diffractometer (Cu K$_{\alpha_{1,2}}$ radiation) was used on crushed single crystals \cite{Jesche2016} and fitted by Rietveld refinement method using Jana 2006 software \cite{Petricek2000}. \\ 

The magnetization measurements, as a function of temperature and magnetic field, were done in a Quantum Design (QD) Magnetic Property Measurement System (MPMS3) SQUID magnetometer in the temperature range 1.8~K $\leq$ T $\leq$ 300~K and magnetic field range $\lvert$H$\rvert$ $\leq$ 7~T.  The magnetization measurements were done on single crystal pieces of 1-3 mg mass, mounted on a quartz sample holder with a small amount of Dow Corning vacuum grease thus ensuring a very small background signal. Temperature and field dependent transport measurements were done in a QD, Physical Property Measurement System for 2~K $\leq$ T $\leq$ 300~K and $\lvert$H$\rvert$ $\leq$ 9~T by using a 3~mA excitation current with a frequency of 17~Hz. The transport measurements were done on thin bars of the single crystals which were shaped by a wire-saw. The contacts for the electrical transport measurements were prepared in a standard four-probe configuration. Four Pt wires of 0.05~mm diameter were attached using Epotek-H20E silver epoxy which was cured at 120$^{\circ}$C for 20 minutes. The contact resistance value was around 4 to 5 $\Omega$. The magnetic field was swept from -9~T to 9~T during the transport measurement. The MR curves were symmetrized to remove the Hall component which, as will be discussed below, can be substantial.\\

\section{RESULTS AND DISCUSSION}
\subsection{X-ray diffraction and crystal structure}
\begin{figure}[t!]
\includegraphics[width=1\linewidth]{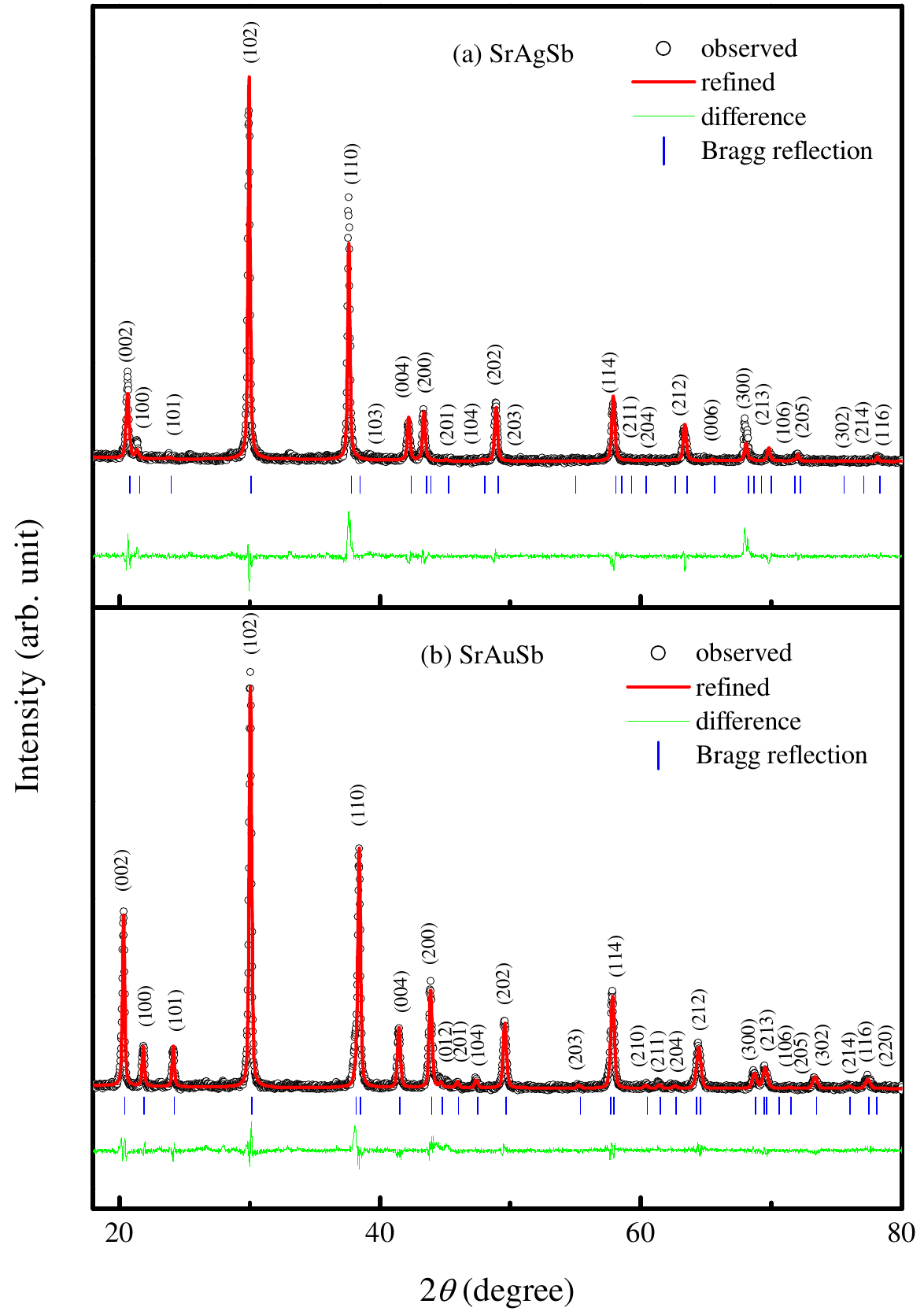}
\centering
\caption{Powder x-ray diffraction pattern of ground single crystal (a) SrAgSb, (b) SrAuSb. The experimental data, fitted curves and residues are shown by black circles, red and green lines, respectively. Blue tick marks represent the Bragg-peak positions corresponding to the \textit{P}6$_{3}$/\textit{mmc} crystal structure.}
\label{XRD1}
\end{figure}
Figure \ref{XRD1} shows the Rietveld refinement of the powder XRD data, fitting with \textit{P}6$_{3}$/\textit{mmc} crystal structure (I$_{cal}$) and difference between measured data and fitting (I$_{diff}$). We have observed Sb-impurity peaks due to residual droplets of flux on the surface for SrAgSb. From the powder XRD data SrAgSb and SrAuSb crystallize in the hexagonal \textit{P}6$_{3}$/\textit{mmc} crystal structure as reported in the literature \cite{Tomuschat1984, Mishra2011}. We first used Le Bail refinement was done which refines only the unit cell parameter and profile broadening function. Next, Rietveld refinement was done so as to gain information about the atomic position. Only lattice parameters, profile broadening function were refined. The fixed Wyckoff sites of Sr, Ag/Au and Sb were used during the refinement. These Wyckoff sites were found to be consistent as reported in literature \cite{Merlo1990}. The inferred crystal structure parameters are listed in table \ref{tab1}. The atomic positions are listed in table \ref{tab2}.\\
\begin{table}[h!]
\begin{center}
\begin{tabular}{ c c c }
 \hspace{0.2 cm}{Compound}\hspace{0.2 cm}&\hspace{0.2 cm}{SrAgSb}\hspace{0.2 cm}&\hspace{0.2 cm}{SrAuSb}\hspace{0.2 cm}\\\hline\hline
$a$ ($\AA$)&\hspace{0.2 cm}4.7628(3)\hspace{0.2 cm}&\hspace{0.2 cm}4.6772(2)\hspace{0.2 cm}\\
$c$ ($\AA$)&\hspace{0.2 cm}8.5343(7)\hspace{0.2 cm}&\hspace{0.2 cm}8.6962(5)\hspace{0.2 cm}\\
$\alpha$ ($^{\circ}$)&\hspace{0.2 cm}90\hspace{0.2 cm}&\hspace{0.2 cm}90\hspace{0.2 cm}\\
$\beta$ ($^{\circ}$)&\hspace{0.2 cm}90\hspace{0.2 cm}&\hspace{0.2 cm}90\hspace{0.2 cm}\\
$\gamma$ ($^{\circ}$)&\hspace{0.2 cm}120\hspace{0.2 cm}&\hspace{0.2 cm}120\hspace{0.2 cm}\\
cell volume ($\AA^{3}$)&\hspace{0.2 cm}167.66(2)\hspace{0.2 cm}&\hspace{0.2 cm}164.75(2)\hspace{0.2 cm}\\\hline
\end{tabular}
\end{center}
\caption{Lattice parameters obtained from Rietveld refinement of powder x-ray diffraction data}
\label{tab1}
\end{table}  

\begin{figure}[t!]
\includegraphics[width=1\linewidth]{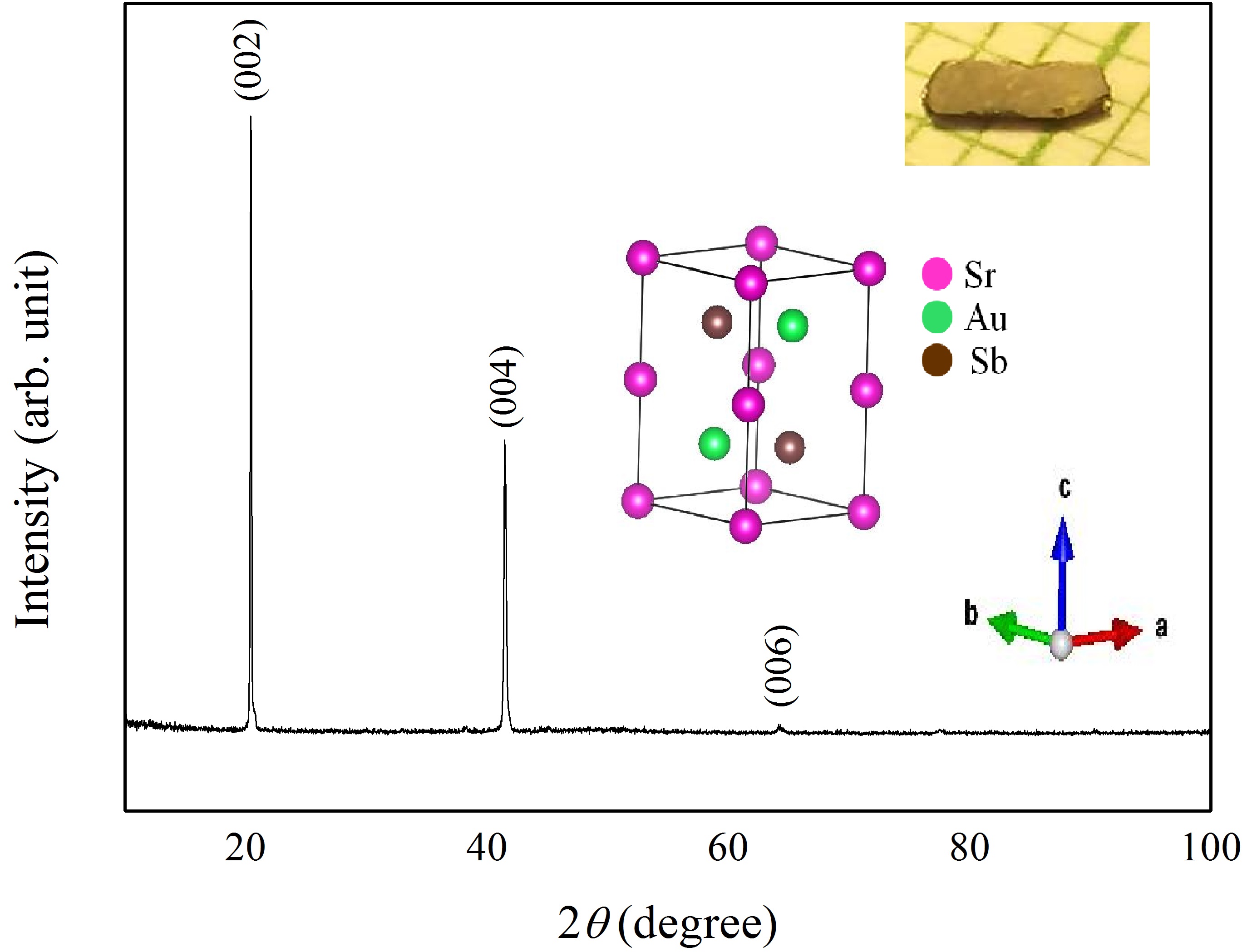}
\centering
\caption{Single crystal X-ray diffraction pattern shows the $\theta$ $-$ 2$\theta$ scan on one SrAuSb single crystal plate for the x-ray incidence angle $\theta$ with the plane perpendicular to the single crystal plate as a representative, identifying the (0 0 $l$) lines. Insets show the plate-like SrAuSb crystal on a mm grid and the image of unit cell of SrAuSb compound.}
\label{XRD2}
\end{figure}

\begin{table}[h!]
\begin{center}
\begin{tabular}{ c c c c c }
 \hspace{0.2 cm}{Atom}\hspace{0.2 cm}&\hspace{0.2 cm}{Wyck}\hspace{0.2 cm}&\hspace{0.2 cm}${x}$\hspace{0.2 cm}&\hspace{0.2 cm}${y}$\hspace{0.2 cm}&\hspace{0.2 cm}${z}$\hspace{0.2 cm}\\\hline\hline
Sr&\hspace{0.2 cm}2$a$\hspace{0.2 cm}&\hspace{0.2 cm}0\hspace{0.2 cm}&\hspace{0.2 cm}0\hspace{0.2 cm}&\hspace{0.2 cm}0\hspace{0.2 cm}\\
Ag/Au&\hspace{0.2 cm}2$d$\hspace{0.2 cm}&\hspace{0.2 cm}1/3\hspace{0.2 cm}&\hspace{0.2 cm}2/3\hspace{0.2 cm}&\hspace{0.2 cm}0.75\hspace{0.2 cm}\\
Sb&\hspace{0.2 cm}2$c$\hspace{0.2 cm}&\hspace{0.2 cm}1/3\hspace{0.2 cm}&\hspace{0.2 cm}2/3\hspace{0.2 cm}&\hspace{0.2 cm}0.25\hspace{0.2 cm}\\\hline
\end{tabular}
\end{center}
\caption{Atomic coordinates of the SrAgSb and SrAuSb structure obtained from Rietveld analysis of the powder x-ray diffraction data}
\label{tab2}
\end{table}
We have also done a $\theta$ $-$ 2$\theta$ scan on a single crystal piece of SrAuSb to determine the crystallographic ab-plane and c-axis of the plate-like crystal \cite{Jesche2016}. When the x-ray beam is incident perpendicular to the surface, only (00\textit{l}) reflections were obtained as shown in figure \ref{XRD2}. The inferred value of lattice parameter \textit{c} is very close to the \textit{c} value of their x-ray powder diffraction. The insets of figure \ref{XRD2} show the plate-like crystal on mm grid and the unit cell of SrAuSb compound drawn by VESTA \cite{Momma2008}. \\

\subsection{Band structure and topological features}
\begin{figure}[h!]
\includegraphics[width=1\linewidth]{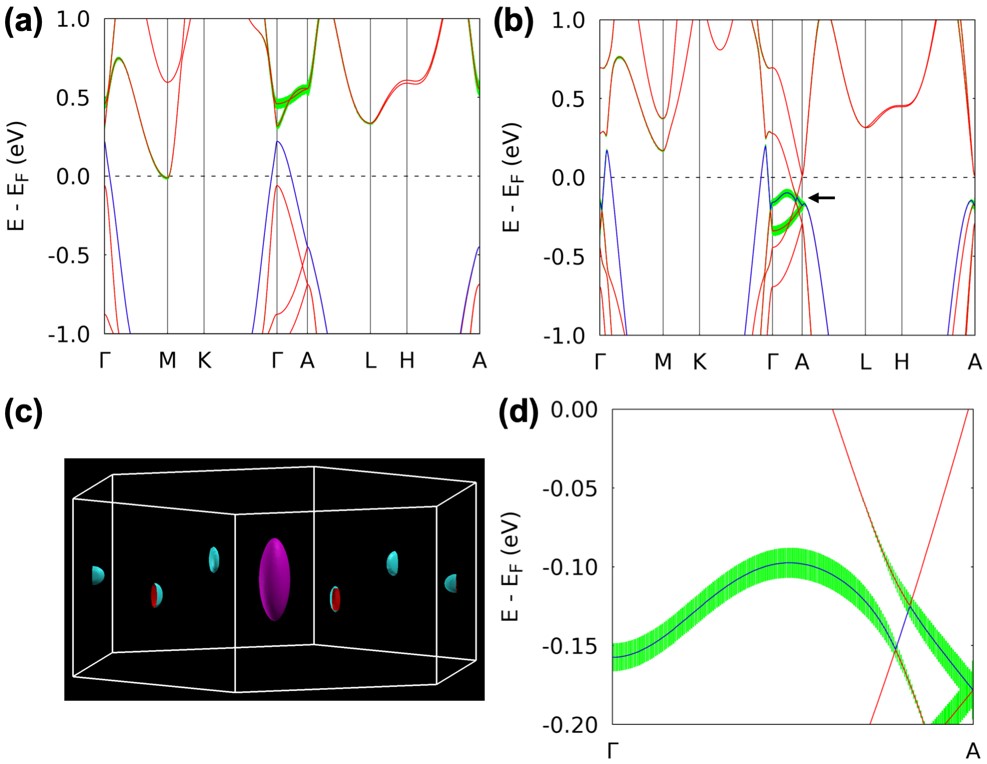}
\centering
\caption{Band structure of (a) SrAgSb and (b) SrAuSb. The top valence band according to band filling is in blue. The green shade stands for the projection on Ag 5$s$ or Au 6$s$ orbitals. The black arrow points to the bulk Dirac point in SrAuSb. (c) 3D Fermi surface for SrAgSb at E$_{F}$ showing the hole pocket at $\Gamma$ and electron pockets at M points. (d) zoom in along $\Gamma-A$ direction for SrAuSb showing the band inversion and the Dirac point at E$_{F}$-0.125 eV.}
\label{bandstructure}
\end{figure}

BeZrSi structure in space group 194 is known to host a variety of compounds with topological non-trivial states \cite{Gibson2015}. To search for topological features in SrAgSb and SrAuSb, we have performed DFT \cite{Hohenberg1964, Kohn1965} band structure calculations. Figures \ref{bandstructure}(a) and (b) are the bulk band structures of SrAgSb and SrAuSb, respectively. SrAgSb is a trivial semi-metal with a band gap between its valence and conduction bands, but without a global band gap. The valence band is mostly derived from Sb 5$p$ orbitals and the conduction band from Sr and Ag 5$s$ orbitals. Specifically, the two conduction bands along $\Gamma$-A direction are derived from Ag 5$s$ orbitals, which are just above the top of valence band. The Fermi surface in figure \ref{bandstructure}(c) shows a hole pocket at $\Gamma$ and six electron pockets at M points. Thus, SrAgSb is a compensated semimetal that can have non-saturating MR \cite{Ali2014, Lifshits1973}. In terms of band structure topology, there is no band inversion across the continuous band gap. Calculations of band topology with symmetry-based indicator \cite{Tang2019, Po2017, Khalaf2018}, elemental band representation \cite{Bradlyn2017} and layer construction \cite{Zhang2019, Song2018} show that  SrAgSb is topologically trivial. Note SrAgSb is classified as a topological insulator in the topological material databases \cite{Tang2019, Zhang2019}. But those calculations were based on the crystal structure with Ag at the Wyckoff 2$a$ (0, 0, 0) position, which is different from published data \cite{Merlo1990} as well as our own XRD data and analysis. The 2$a$ position is larger than 2$b$ and 2$c$ positions and should accommodate the larger Sr atom. DFT total energy calculations confirm that the structure with Sr at 2$a$ position is a few eV per formula unit more stable than the one with Ag at 2$a$ position.\\

When replacing Ag with Au for SrAuSb (see figure \ref{bandstructure}(b)), due to the much stronger relativistic effect of Au than Ag, the Au 6$s$ derived band along the $\Gamma-A$ direction is pulled down to lower energy and giving rise to a band inversion. Such a band inversion brings non-trivial topological behavior. The crossing point from the band inversion along the $\Gamma-A$ direction is protected by the 3-fold rotational symmetry thus this is a Dirac point, which is marked by an arrow and shown in detail in figure \ref{bandstructure}(d). The Dirac point is at E$_{F}$-0.125 eV with linearly dispersed valence band reaching the E$_{F}$. Such topological state with Dirac point near E$_{F}$ is known to also give non-saturating MR due to the spilitting of the Dirac point into Weyl points under magnetic field \cite{Armitage2018}.  

\subsection{Magnetic and Transport Properties}
\begin{figure}[t!]
\includegraphics[width=1\linewidth]{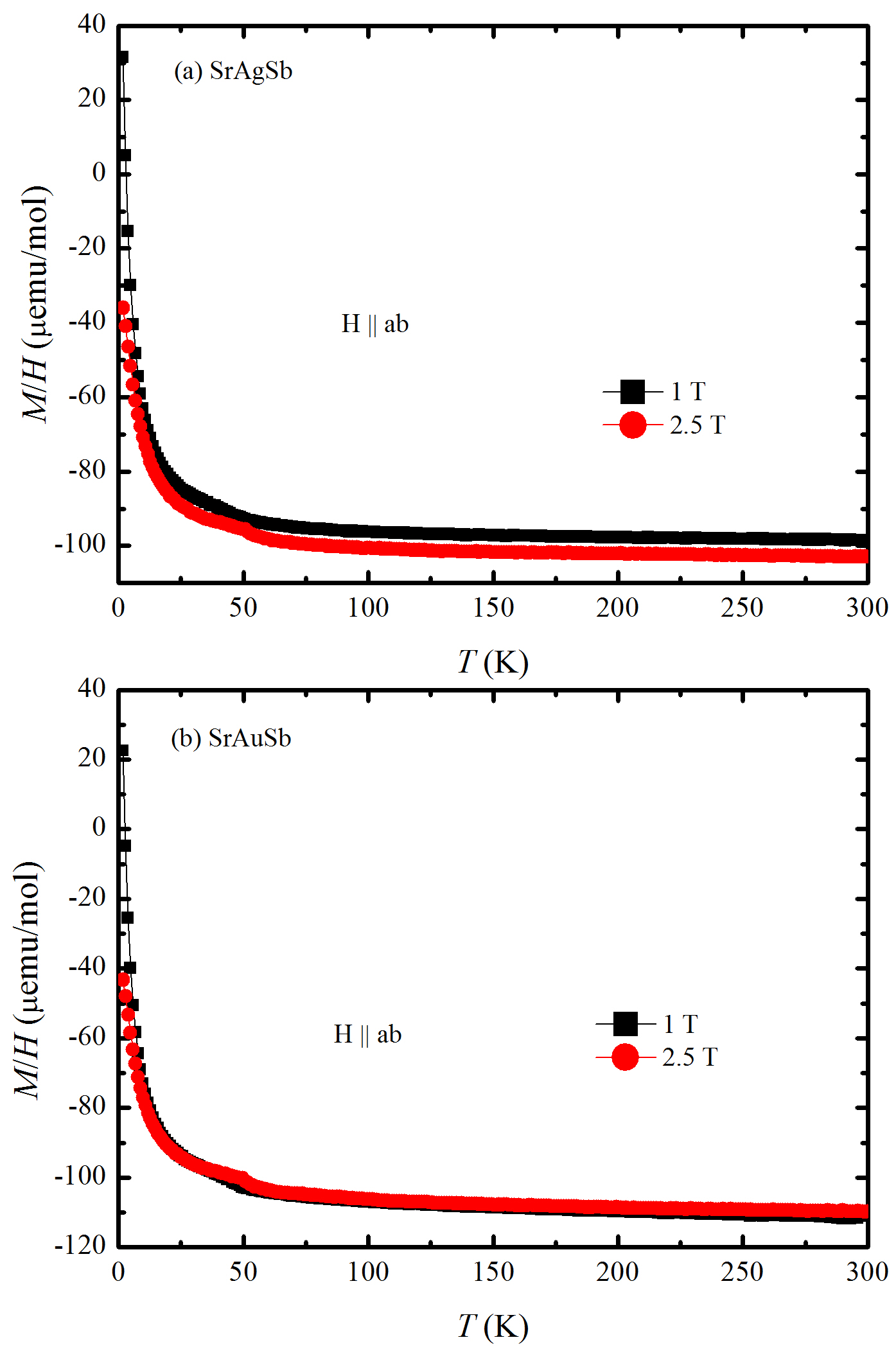}
\centering
\caption{Temperature dependent inverse magnetic susceptibility for H$\parallel$ab ($\chi_{ab}$), at two different magnetic fields of 1~T and 2.5~T for (a) SrAgSb and (b) SrAuSb.}
\label{Magnetization3}
\end{figure}
Magnetization and transport measurements were done on single crystals of SrAgSb and SrAuSb. Figure \ref{Magnetization3} shows the dc susceptibility data for SrAgSb and SrAuSb at two different magnetic field values of 1~T and 2.5~T. The dc susceptibility data shows that both compounds are diamagnetic. However, the low temperature Curie tails indicate that a small amount of magnetic impurity is also likely present.\\ 
\begin{figure}[t!]
\includegraphics[width=0.9\linewidth]{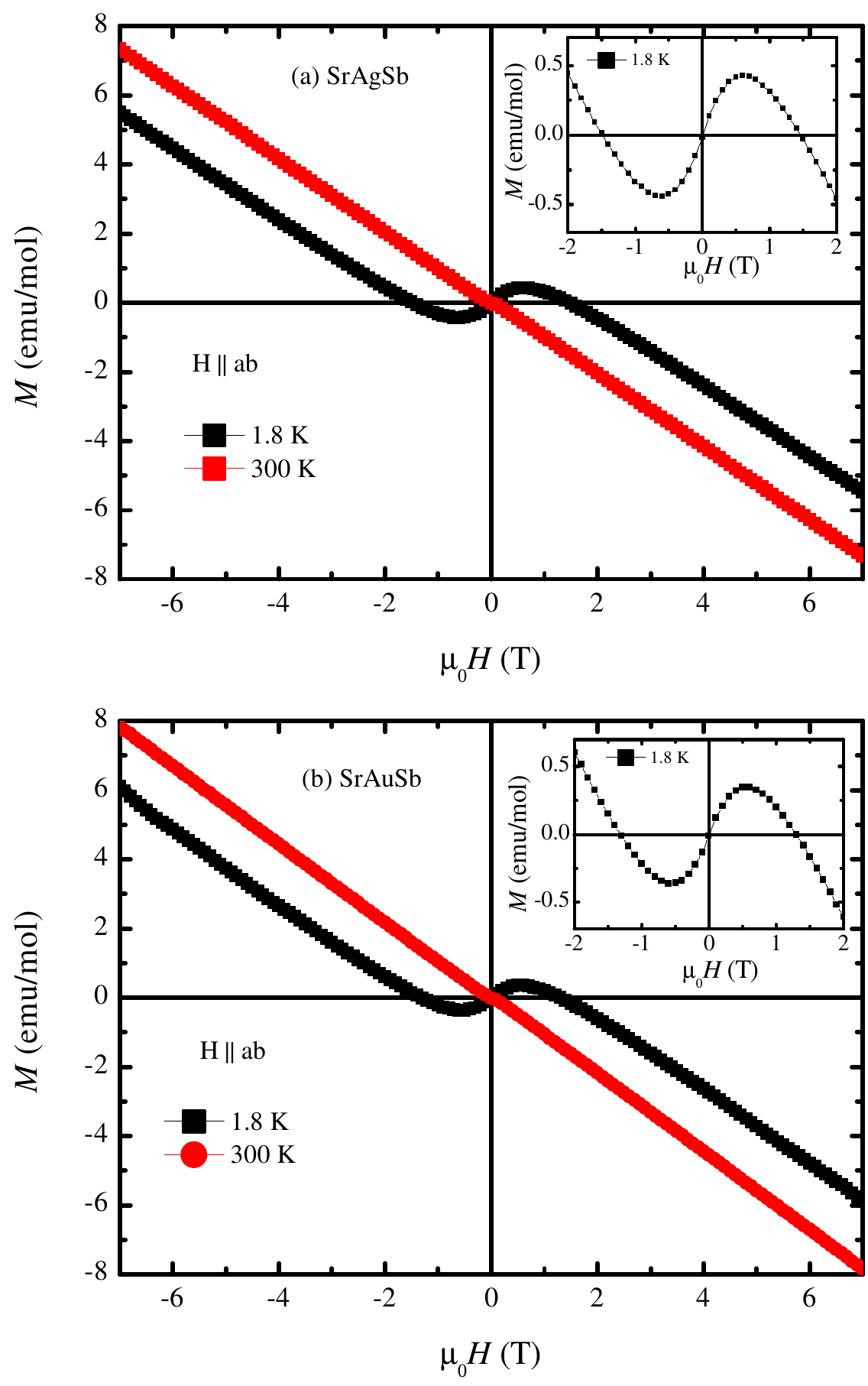}
\centering
\caption{Field dependent magnetization measurements $M$ vs $H$ in the magnetic field range of -7~T to +7~T by applying H$\parallel$ab for (a) SrAgSb and (b) SrAuSb at two different temperatures 1.8~K and 300~K. Inset shows the corresponding saturated magnetic moment on an expanded scale.}
\label{Magnetization4}
\end{figure}

Field dependent magnetization measurements were also measured at two different temperatures, 1.8~K and 300~K, as shown in figure \ref{Magnetization4}(a) and (b) for SrAgSb and SrAuSb, respectively. The insets to figure \ref{Magnetization4}(a) and (b) show the 1.8~K data on an expanded scale in the field range of -2~T to +2~T. The small paramagnetic saturation in the low temperature $M(H)$ data strongly suggest that both single crystals of SrAgSb and SrAuSb have a small amount of magnetic impurity. Moreover, the insets of figure \ref{Magnetization4}(a) and (b) show that the saturated magnetic moment is similar for both compounds which suggest that the magnetic impurity either seems to come from Sr or Sb.\\
\begin{figure}[t!]
\includegraphics[width=1\linewidth]{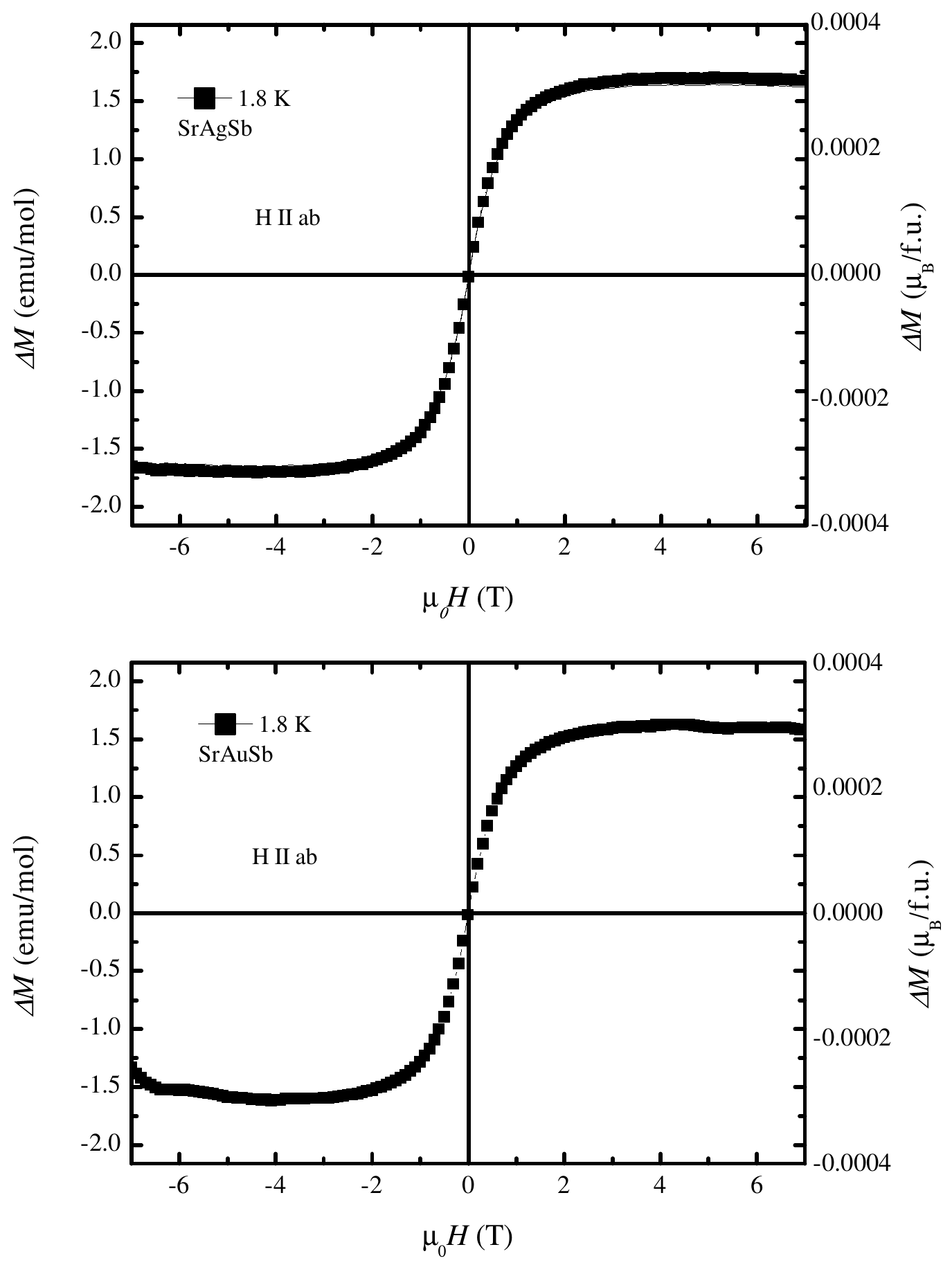}
\centering
\caption{Saturated magnetic moment extracted from the $M(H)$ data at 1.8~K for H$\parallel$ab of both compounds (a) SrAgSb and (b) SrAuSb. Left scale shows the magnetic moment in emu/mol and right scale in $\mu_{B}$/f.u.}
\label{Saturated Moment}
\end{figure}

Given that the $H$ $>$ 2~T, $M(H)$ data have the same slope for T = 1.8~K and T = 300~K, and given that the $\chi$(T) data in figure \ref{Magnetization3} is essentially temperature independent for T $>$ 100~K, we can use the T = 300~K $M(H)$ data as an approximation of the T = 1.8~K $M(H)$ of a pure sample. Based on this, figures \ref{Saturated Moment} show the ``$\Delta$ $M(H)$ = $M$(1.8~K) - $M$(300~K)'' data of SrAgSb and SrAuSb, respectively. As can be seen, when expressed in units of $\mu_{B}$/f.u., this is a very small contribution. If for example, we associate this $\Delta$ $M$ with a hypothetical Eu$^{2+}$ impurity, we would infer approximately 1 mole of Eu$^{2+}$ in 10$^{5}$ mole of Sr. A similarly low impurity level can be extracted from the small Curie tails shown in figure \ref{Magnetization3}.\\  

It should be noted that similar signatures in $M(H)$ data have been given much more exotic interpretations, even claimed to be associated with novel topologies \cite{Pariari2017, Zhao2014}. Such claims, should, perhaps, be examined in light of similar possible explanations of small amounts of magnetic impurities (be them paramagnetic, as in the case, or even small ferromagnetic phases). In the absence of other, clarifying data, the Ptolemaic percept stated as: ``We consider it a good principle to explain the phenomena by the simplest hypothesis possible'' \cite{Franklin2001} to apply in our case.\\
\begin{figure}[t!]
\includegraphics[width=1\linewidth]{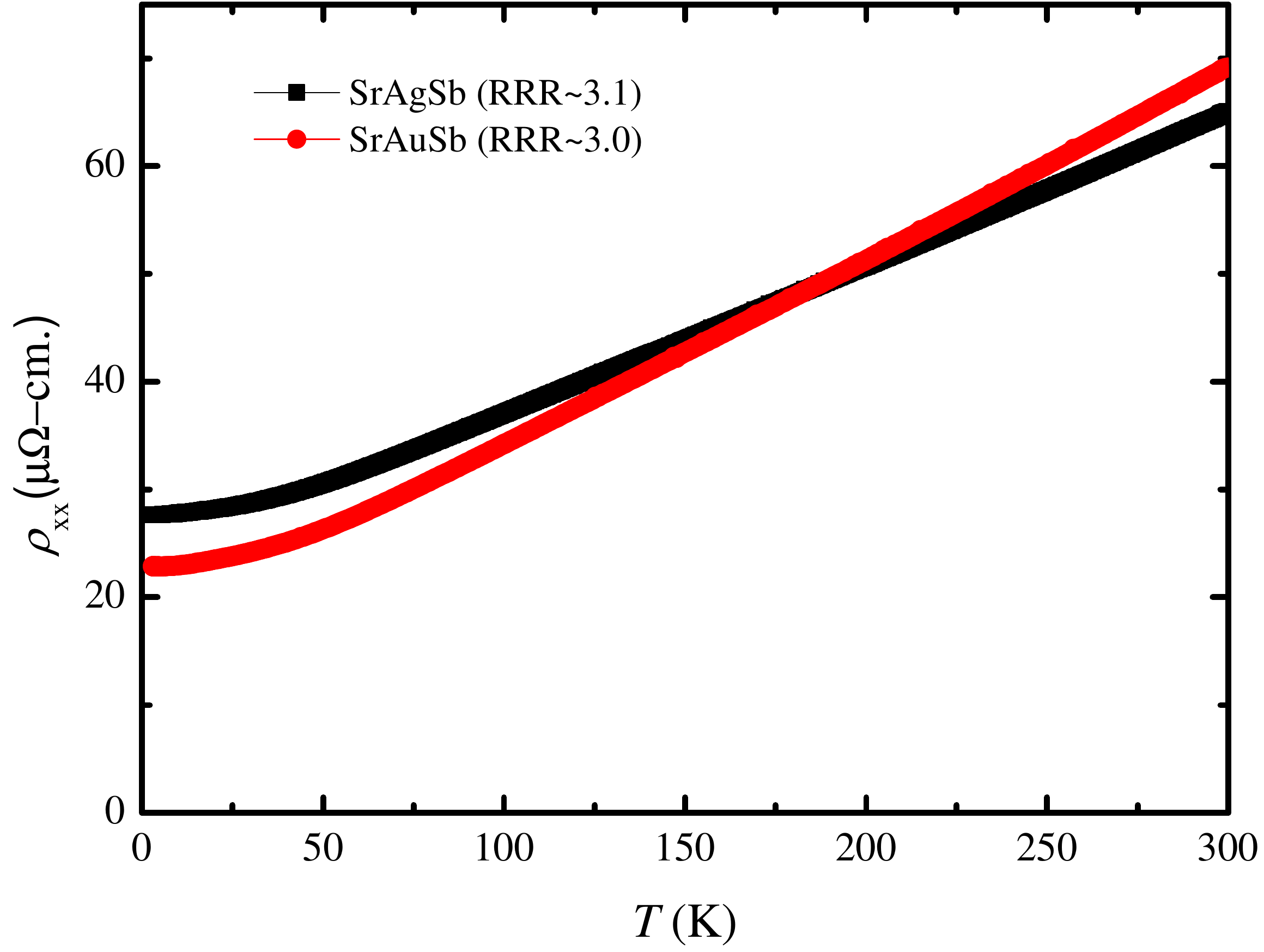}
\centering
\caption{The temperature dependent resistivity in the temperature range from 3~K to 300~K shown by black filled rectangles for SrAgSb and red filled circles for SrAuSb.}
\label{Resistivity2}
\end{figure}

Figure \ref{Resistivity2} shows the resistivity as a function of temperature of SrAgSb (black filled rectangle) and SrAuSb (red filled circle) samples. Multiple samples of each compound were measured; in all cases RRR values ranged from 2.4 $-$ 3.1. SrAgSb and SrAuSb samples exhibit very similar temperature dependence, initially decreasing with decrease of temperature and then showing weak temperature dependence below 30~K.The similar RRR-values suggest similar crystalline quality of the grown single crystals as well. These values of RRR are significantly lower than observed in comparison to other Dirac and Weyl semimetals candidates as well as other Sr based compounds \cite{Schoop2016, Singha2017} and comparable to SrZnBi$_{2}$ \cite{Wang2017}, indicate and large scattering  possibly on defects and dislocations, in our samples.\\
\begin{figure}[t!]
\includegraphics[width=0.9\linewidth]{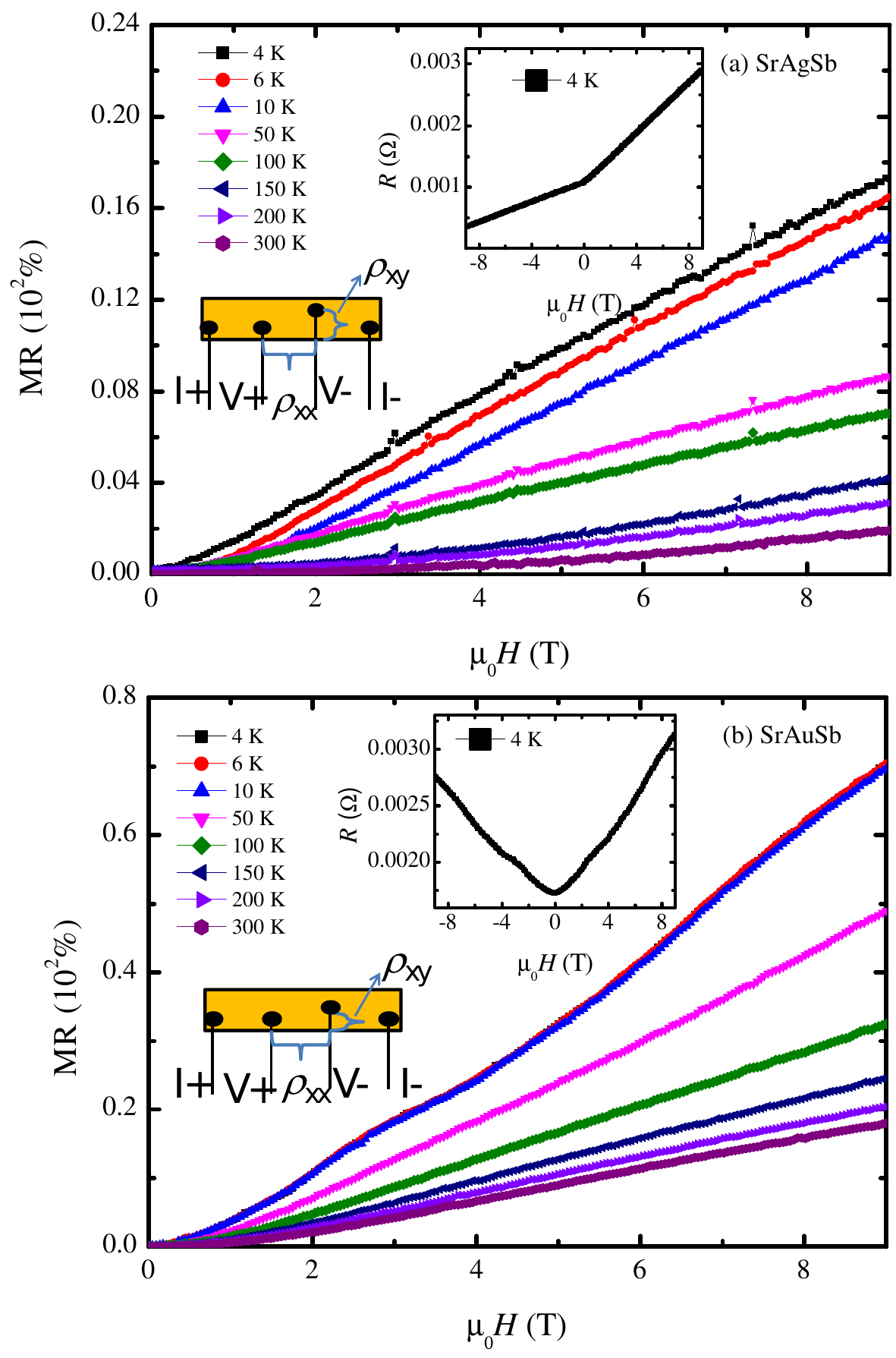}
\centering
\caption{Transverse MR by applying the magnetic field H$\parallel$c at different representative temperatures of (a) SrAgSb and (b) SrAuSb. Insets show their corresponding resistance at 4~K in the magnetic field range of -9~T to +9~T and the schematic of the sample with the four probe contacts}
\label{Resistivity3}
\end{figure}

Recently, large non-saturating MR was reported in several topological non-trivial compounds \cite{Shekhar2015, Ali2014, Schoop2016}. To calculate the MR in SrAgSb and SrAuSb, we measured the resistance as a function of the magnetic field from -9~T to 9~T in the temperature range from 4 to 300~K. The insets to figure \ref{Resistivity3}(a) and (b) show the resistance data at 4~K for SrAgSb and SrAuSb with their corresponding schematics of the sample for the MR measurement. The resistance of SrAgSb and SrAuSb in inset to figure \ref{Resistivity3}(a) and (b) show qualitatively similar, but quantitatively very different behavior. Such differences can occur, if the voltage contacts are slightly displaced from parallel configuration then in addition to the $\rho_{xx}$ component, one can get an additional Hall contribution $\rho_{xy}$ from the perpendicular part as indicated by the schematics in insets to figure \ref{Resistivity3}(a) and (b). Due to an apparently larger misalignment between the two voltage contacts, we got a correspondingly larger Hall contribution for the SrAgSb sample in comparison to the SrAuSb sample. Whereas SrAuSb shows an almost symmetric behavior of the resistance in the positive and negative magnetic field, SrAgSb manifests a highly asymmetric behavior. These asymmetries are associated with the $\rho_{xy}$ (Hall) term.\\ 

To remove the $\rho_{xy}$ term from the $\rho_{xx}$ MR data, the resistivity should be measure in both negative and positive field direction and then it can be symmetrized by using the formula ($\rho_{(B)}$ + $\rho_{(-B)}$)/2. It is possible to exclude one contribution from the other because $\rho_{xx}$ is symmetric, and $\rho_{xy}$ is antisymmetric as a function of magnetic field. Thus,  from the symmetrization of resistivity in the negative and positive field range, the $\rho_{xy}$ component can be removed from the $\rho_{xx}$ component.\\ 

Also, figures \ref{Resistivity3}(a) and (b) show the symmetrized, transverse MR i.e. MR under transverse electric and magnetic fields in a series of isothermal $R(H)$ data sets for SrAgSb and SrAuSb compounds, by applying the current along the hexagonal plane $ab$ and magnetic field along the $c$ axis. At 4~K and 9~T, MR values of  17\% and 70\% were obtained without any signature of saturation for SrAgSb and SrAuSb respectively, which is comparable to CaMnBi$_{2}$ \cite{Wang2012}. SrAuSb show approximately four times larger MR in comparison to SrAgSb. However, this value of MR is lower than observed in several topological semimetals \cite{Xiang2015}. \\

\section{CONCLUSION}
In conclusion, we have shown that SrAgSb and SrAuSb compounds crystallize in hexagonal structure with the centrosymmetric space group \textit{P}6$_{3}$/mmc. The band structure calculations were done on the basis of experimentally determined Wyckoff sites which show that SrAgSb is compensated semimetal, whereas SrAuSb is Dirac semimetal due to large spin-orbit interaction of Au in comparison to Ag atom. Magnetic measurement of both SrAgSb and SrAuSb show diamagnetic behavior with small magnetic impurities giving a low temperature Curie tail and small saturated moments. The temperature dependent resistivity of both compounds is metal-like with low RRR-values suggesting substantial impurity scattering. The magnetization and resistivity data show that there is no phase transition in either compound. At 9 T and 4 K, a transverse MR of approximately 17\% for SrAgSb and 70\% for SrAuSb has been observed without any signature of saturation. This non-saturating MR in non magnetic Sr based compounds can be understood from the band structure calculation. Note that MR in SrAuSb is approximately four time larger and this might be associated with topological change but futher, detailed, temperature dependent magnetotransport, possibly on better samples would be needed. \\

\section*{Acknowledgments}
We thank Elena Gati for her great assistance in the measurements and discussion and Dominic Ryan for the help in the XRD analysis. Work at the Ames Laboratory was supported by the U.S. Department of Energy, Office of Science, Basic Energy Sciences, Material Sciences and Engineering Division. The Ames laboratory is supported for the U.S. Department of Energy by Iowa State University under Contract No. DEAC0207CH11358. P. D. was supported by the Center for the Advancement of Topological Semimetals, an Energy Frontier Research Center funded by the U.S. DOE, Office of Basic Energy Sciences.

\bibliography{mybibfile}

\end{document}